\def\rfr#1{eq. (\ref{#1})}

\def\Rfr#1{Eq. (\ref{#1})}

\def\dert#1#2{\frac{{{d}}{#1}}{{{d}}{#2}}}              % derivate parziali e totali prima e seconda

\def\bar{\begin{eqnarray}}
\def\ear{\end{eqnarray}}
\def\bb{\bibitem}
\def\eqi{\begin{equation}}
\def\eqf{\end{equation}}
\def\eqia{\begin{eqnarray}}
\def\eqfa{\end{eqnarray}}
\def\rp#1#2{{#1\over#2}}

\def\lb#1{\label{#1}}

\def\oc2{$\mathcal{O}(c^{-2})$}

\documentclass[11pt]{article}
\usepackage{amsmath,amsthm,amscd,amssymb}
\usepackage{latexsym}
\usepackage{graphicx,epsfig}

\begin{document}

\noindent{\bf \LARGE{On a recently proposed metric linear
extension of general relativity to explain the Pioneer anomaly}}
\\
\\
\\
{Lorenzo Iorio}\\
{\it Viale Unit$\grave{a}$ di Italia 68, 70125\\Bari, Italy
\\tel./fax 0039 080 5443144
\\e-mail: lorenzo.iorio@libero.it}

\begin{abstract}
Recently, Jaekel and Reynaud put forth a  metric linear extension
of general relativity which, in the intentions of its proponents,
would be able, among other things, to provide a gravitational
mechanism for explaining the Pioneer anomaly without contradicting
either the equivalence principle or what we know about the
planetary motions. In this paper we perform an independent test of
such an hypothesis by showing that the planets' orbits are, in
fact, affected by the suggested mechanism as well, and comparing
the resulting effects with the latest observational
determinations. It turns out that the predicted perihelion
precessions, expressed in terms of an adjustable free parameter
$\zeta_{\rm P}M$ set equal to the value used to reproduce the
magnitude of the Pioneer anomalous acceleration, are quite
different from the observationally determined extra-advances of
such Keplerian element for the inner planets. Conversely, the
values obtained for $\zeta_{\rm P}M$ from the determined
perihelion extra-rates of the inner planets turn out to be in
disagreement with the value which would be required to accommodate
the Pioneer anomaly. As a consequence, the suggested explanation
 for the Pioneer anomaly, based on the
assumption that $\zeta_{\rm P}$ is constant throughout the Solar
System,  should be rejected, at least in its present form.
\end{abstract}



\section{Introduction}
In order to find an explanation of gravitational origin for the
anomalous acceleration of about $(8.74\pm 1.33)\times 10^{-10}$ m
s$^{-2}$  experienced by the Pionner 10/11 spacecraft after they
passed the threshold of 20 AU (Anderson et al. 1998; 2002), Jaekel
and Reynaud (2005a; 2005b) proposed to use a suitable metric
linear extension of General Relativity with two potentials
$\Phi_{\rm N}$ and $\Phi_{\rm P}$. In the gauge convention of the
PPN formalism its space-time line element, written in isotropic
spherical coordinates, is (Jaekel and Reynaud 2005b) \eqi
ds^2=g_{00}c^2dt^2+g_{rr}[dr^2+r^2(d\theta^2+\sin^2\theta
d\phi^2)],\eqf with
\begin{equation}\left\{\begin{array}{lll}g_{00}=1+2\Phi_{\rm
N},\\\\
g_{rr}=-1+2\Phi_{\rm N}-2\Phi_{\rm
P}.\lb{poti}\end{array}\right.\end{equation} In order to
accommodate the Pioneer anomaly  the following simple model \eqi
\Phi_j(r)=-\rp{G_j M }{c^2 r}+\rp{\zeta_j M r}{c^2}, j={\rm N,
P},\lb{potii}\eqf has been used (Jaekel and Reynaud 2005a; 2005b).
It is determined by four constants: the Newtonian constant $G_{\rm
N}$ and the three small parameters $G_{\rm P}, \zeta_{\rm N }$ and
$\zeta_{\rm P}$ which measure the deviation from general
relativity. In the intentions of Jaekel and Reynaud, their theory
should be able to explain the occurrence of the Pioneer anomaly
 a) without violating either the existing constraints from the
planetary motions b) or the equivalence principle. The latter goal
is ensured by the metric character of the proposed extension of
general relativity. In regard to a), they first focus their
attention to the modification of the Newtonian potential. By using
the orbits of Mars and the Earth they get an upper bound $|\zeta
_{\rm N}M|\simeq 5\times 10^{-13}$ m s$^{-2}$ (Jaekel and Reynaud
2005b) which excludes that $\zeta_{\rm N}M r/c^2$ is capable to
account for the anomalous Pioneer acceleration. The key point of
their line of reasoning in explaining the Pioneer anomaly without
contradicting our knowledge of the planetary orbits consists in
considering from the simple expression of \rfr{potii} for
$\Phi_{\rm P}$ the following extra-kinetic radial
acceleration\footnote{The contribution of $G_{\rm P}$ is found to
be negligible.} \eqi A_{\rm P }=2\zeta_{\rm P}M\rp{v^2_r}{c^2},
\lb{accel}\eqf where $v_r$ is the radial component of the velocity
of the moving body, in identifying it with the source of the
Pioneer anomalous acceleration by getting\footnote{The almost
constant value $v_r=1.2\times 10^4$ m s$^{-1}$ has been used for
both the Pioneer spacecraft.} $\zeta_{\rm P}M=0.25$ m s$^{-2}$ and
in claiming that \rfr{accel} cannot affect the planetary motions
because almost circular. Conscious of the fact that independent
tests are required to support their hypothesis and since no
accurate and reliable data from other spacecraft are available to
this aim, Jaekel and Reynaud (2005b) propose to perform light
deflection measurements because $\Phi_{\rm P}$ affects the motion
of electromagnetic waves as well. A re-analysis of the Cassini
(Bertotti et al. 2003) data is suggested (Jaekel and Reynaud
2005b).

In this paper we will show that, in fact, the extra-kinetic
acceleration of \rfr{accel} does also affect the orbital motions
of the planets in such a way that it is possible to compare the
resulting features of motion with the latest data from planetary
ephemerides, thus performing right now a clean and independent
test of the hypothesis that \rfr{accel} is able to accommodate the
Pioneer anomaly. We will also discuss the feasibility of the
proposed light deflection measurements in view of the results
obtained from the perihelia test.
\section{The orbital effects of the kinetic acceleration and comparison with the latest data}
The Russian astronomer E.V. Pitjeva  has recently processed almost
one century of data of all types in the effort of continuously
improving the EPM2004 planetary ephemerides. Among other things,
she also determined residual advances of the perihelia $\omega$ of
the inner (Pitjeva 2005) and outer (Pitjeva 2006) Solar System
planets as fit-for parameters of a global solution in which she
contrasted, in a least-square way, the observations to their
predicted values computed with a complete set of dynamical force
models including all the known Newtonian and Einsteinian features
of motion. As a consequence, any unmodelled force, as it would be
the case for a Pioneer-like one if present in Nature, is entirely
accounted for by the so-obtained residual perihelia advances.

In order to make a direct comparison with them, we will now
analytically work out the secular, i.e. averaged over one orbital
revolution, effects induced by the extra-kinetic acceleration of
\rfr{accel} on the pericentre of a test particle. To this aim, we
will treat \rfr{accel} as a small perturbation of the Newtonian
monopole. In order to justify this assumption, we will first
evaluate the average of \rfr{accel} and, then, we will compare it
with the the Newtonian  mean accelerations throughout the Solar
System. To this aim, we must evaluate \rfr{accel} onto an
unperturbed Keplerian ellipse by means of \eqi v_r=\rp{nae\sin
f}{\sqrt{1-e^2}},\lb{vrad}\eqf where $a$ is the semimajor axis,
$e$ is the eccentricity, $n=\sqrt{GM/a^3}$ is the (unperturbed)
Keplerian mean motion and $f$ is the true anomaly. Subsequently,
the average over one orbital period $P=2\pi/n$ has to  be
performed. It is useful to adopt the eccentric anomaly $E$ by
means of the relations
\begin{equation}\left\{\begin{array}{lll}
dt=\rp{(1-e\cos E)}{n}dE
,\\\\
\cos f=\rp{\cos E-e}{1-e\cos E
},\\\\
\sin f=\rp{\sin E\sqrt{1-e^2}}{1-e\cos E}.
\lb{eccen}\end{array}\right.\end{equation}
By using \eqi\int_0^{2\pi}\rp{\sin^2 E}{1-e\cos E }
dE=\rp{2\pi}{e^2}\left(1-{\sqrt{1-e^2}}\right),\eqf we get
\eqi\left\langle A_{\rm P}\right\rangle=\rp{2\zeta_{\rm P}M n^2
a^2}{c^2}\left(1-{\sqrt{1-e^2}}\right).\lb{aveap}\eqf \Rfr{aveap}
can, now, be compared with \eqi \left\langle A_{\rm
N}\right\rangle=\rp{GM}{a^2\sqrt{1-e^2}},\lb{chepalle}\eqf The
results are in Table \ref{scassapalle}
{\small\begin{table}\caption{ Average Pioneer and Newtonian
accelerations for the Solar System planets, in m s$^{-2}$. For
$\left\langle A_{\rm P} \right\rangle$ the expression of
\rfr{aveap} has been used with $\zeta_{\rm P}M=0.25$ m s$^{-2}$.
}\label{scassapalle}

\begin{tabular}{lll} \noalign{\hrule height 1.5pt}

Planet  & $\left\langle A_{\rm P} \right\rangle$ & $\left\langle A_{\rm N} \right\rangle$\\
Mercury & $2\times 10^{-10}$   & $4\times 10^{-2}$    \\
Venus   & $1\times 10^{-13}$   & $1\times 10^{-2}$    \\
Earth   & $6\times 10^{-13}$   & $6\times 10^{-3}$    \\
Mars    & $1\times 10^{-11}$   & $2\times 10^{-3}$    \\
Jupiter & $1\times 10^{-12}$   & $2\times 10^{-4}$    \\
Saturn  & $8\times 10^{-13}$   & $6\times 10^{-5}$    \\
Uranus  & $2\times 10^{-13}$   & $1\times 10^{-5}$    \\
Neptune & $6\times 10^{-15}$   & $6\times 10^{-6}$    \\
Pluto   & $4\times 10^{-12}$   & $4\times 10^{-6}$    \\
\hline

\noalign{\hrule height 1.5pt}
\end{tabular}

\end{table}}
From it it clearly turns out that the use of the  perturbative
scheme is quite adequate for our purposes. The Gauss equation for
the variation of $\omega$ under the action of an entirely radial
perturbing acceleration $A_r$ is \eqi\dert\omega t
=-\rp{\sqrt{1-e^2}}{nae}A_r\cos f.\lb{gaus}\eqf After being
evaluated onto the unperturbed Keplerian ellipse by using
\rfr{vrad}, \rfr{accel} must be inserted into \rfr{gaus}; then,
the average over one orbital period has to be taken.
By means of \eqi \int_0^{2\pi}\rp{\sin^2 E(\cos E-e)}{(1-e\cos
E)^2}dE = \rp{2\pi}{e^3}\left(-2+e^2+2\sqrt{1-e^2}\right),\eqf it
is possible to obtain \eqi\dert\omega t=-\rp{2\zeta_{\rm
P}Mna\sqrt{1-e^2}}{c^2
e^2}\left(-2+e^2+2\sqrt{1-e^2}\right).\lb{pippo}\eqf Note that
\rfr{pippo} is an exact result. It may be interesting to note that
the rates for the semimajor axis and the eccentricity turn out to
be zero; it is not so for the mean anomaly $\mathcal{M}$, but no
observational determinations exist for its extra-rate. We will now
use \rfr{pippo} and $\zeta_{\rm P}M=0.25$ m s$^{-2}$, which has
been derived from \rfr{accel} by imposing that it is the source of
the anomalous Pioneer acceleration, to calculate the perihelion
rates of the inner\footnote{The extra-perihelion rates of the
outer planets (Jupiter, Saturn, Uranus) have recently been
determined in a very preliminary way (Pitjeva 2006); it turns out
that the realistic uncertainties are still  so large that they
cannot be used for a meaningful comparison.} planets of the Solar
System for which estimates of their extra-advances accurate enough
for our purposes exist (Pitjeva 2005). The results are summarized
in Table \ref{tavola}
{\small\begin{table}\caption{ (P): predicted extra-precessions  of
the longitudes of  perihelia of the inner planets, in arcseconds
per century, by using \rfr{pippo} and $\zeta_{\rm P}M=0.25$ m
s$^{-2}$. (D): determined extra-precessions  of the longitudes of
perihelia of the inner planets, in arcseconds per century. Data
taken from Table 3 of (Pitjeva 2005). It is important to note that
the quoted uncertainties are not the mere formal, statistical
errors but are realistic in the sense that they were obtained from
comparison of many different solutions with different sets of
parameters and observations (Pitjeva, private communication 2005).
}\label{tavola}

\begin{tabular}{lllll} \noalign{\hrule height 1.5pt}

 & Mercury & Venus  & Earth & Mars\\
(P) & 1.8323 & 0.001 & 0.0075 & 0.1906 \\
(D) & $-0.0036\pm 0.0050$ & $0.53\pm 0.30$ & $-0.0002\pm 0.0004$ & $0.0001\pm 0.0005$\\

\hline

\noalign{\hrule height 1.5pt}
\end{tabular}

\end{table}}

{\small\begin{table}\caption{ Values of $\zeta_{\rm P}M $, in m
s$^{-2}$, obtained from the determined extra-advances of perihelia
(Pitjeva 2005). After discarding the value for Venus, the weighted
mean for the other planets yields $\zeta_{\rm P}M =-0.0001\pm
0.0004$ m s$^{-2}$. The Pioneer anomalous acceleration is,
instead, reproduced for $\zeta_{\rm P}M=0.25$ m s$^{-2}$.
}\label{tavola2}

\begin{tabular}{lllll} \noalign{\hrule height 1.5pt}

 & Mercury & Venus  & Earth & Mars\\
$\zeta_{\rm P}M $ & $-0.0005\pm 0.0007$ & $91\pm 51$ & $-0.006\pm 0.013$ & $0.0001\pm 0.0006$\\
\hline

\noalign{\hrule height 1.5pt}
\end{tabular}

\end{table}}

It clearly turns out that the determined extra-advances of
perihelia are quite different from the values predicted in the
hypothesis that \rfr{accel} can explain the Pioneer anomaly. In
Table \ref{tavola2} we show the values of $\zeta_{\rm P}M$ which
can be obtained from the determined extra-advances of perihelia
(Pitjeva 2005); as can be noted, all of them are far from the
value which would be required to obtain the correct magnitude of
the anomalous Pioneer acceleration. The experimental intervals
obtained from Mercury, the Earth and Mars are compatible each
other; Venus, instead, yields values not in agreement with them.
This fact can be explained by noting that its perihelion is a bad
observable due to its low eccentricity ($e_{\rm Venus}=0.00677$).
By applying the Chauvenet criterion we reject the value obtained
from the Venus perihelion since it lies at almost 2$\sigma$ from
the mean value of the distribution of Table \ref{tavola2}. The
weighted mean for Mercury, the Earth and Mars is, thus,
$\left<\zeta_{\rm P}M\right>_{\rm w}=-0.0001$ m s$^{-2}$ with a
variance, obtained from $1/\sigma^2=\sum_i(1/\sigma_i^2)$, of
0.0004 m s$^{-2}$.

An analysis involving the perihelia of Mars only can be found in
(Jaekel and Reynaud 2006). In it Jaekel and Reynaud present a
nonlinear generalization of their model, and an explicit
approximate expression of the perihelion rate different from
\rfr{pippo} can be found; it\footnote{More precisely, in (Jaekel
and Reynaud 2006) an explicit expression for the adimensional
perihelion shift after one orbital period, in units of $2\pi$,
i.e. $(\dot\omega P)/2\pi$, can be found; a direct comparison with
our results can be done simply by multiplying their formula by $n$
and making the conversion from s$^{-1}$ to arcseconds per
century.} is calculated with $\zeta_{\rm P}M=0.25$ m s$^{-2}$
yielding a value for the Martian perihelion advance which is about
one half of  our value in Table \ref{tavola}. Even in this case,
the results by Pitjeva (2005) for Mars would rule out the
hypothesis that the Pioneer anomaly can be explained by the
proposed nonlinear model. By the way, in (Jaekel and Reynaud 2006)
no explicit comparison with published or publicly available data
is present.

All the previous considerations are based on the simple model of
\rfr{potii}, with $\zeta_{\rm P}$ constant over the whole range of
distances from the radius of the Sun to the size of the Solar
System. Jaekel and Reynaud (2005a; 2005b; 2006), in fact, leave
generically open the possibility that, instead, $\zeta_{\rm P}$
may vary with distance across the Solar System, but neither
specific empirical or theoretical justifications for such a
behavior are given nor any explicit functional dependence for
$\zeta_{\rm P}(r)$ is introduced.
\section{The deflection of light}
The results for $\zeta_{\rm P}M$ from the determined extra-rates
of the perihelia of the inner planets allow us to safely examine
the light deflection measurements originally proposed by Jaekel
and Reynaud as independent tests of their theory; indeed, the
values of Table \ref{tavola2} certainly apply to the light grazing
the Sun, also in the case of an hypothetical variation of
$\zeta_{\rm P}(r)$ with distance. In (Jaekel and Reynaud 2005a)
they found the following approximate expression for the deflection
angle induced by $\zeta_{\rm P}$ \eqi \psi_{\rm
P}=-\rp{2\zeta_{\rm P}M\rho}{c^2}L,\lb{def}\eqf where $\rho$ is
the impact parameter and $L$ is a factor of order of unity which
depends logarithmically on $\rho$ and on the distances of the
emitter and receiver to the Sun. For $\rho=R_{\odot}$, $L\sim 1$,
and $\zeta_{\rm P}M=-0.0001$ m s$^{-2}$, \rfr{def} yields a
deflection of only -0.3 microarcseconds, which can be translated
into an equivalent accuracy of about $2\times 10^{-7}$ in
measuring the PPN parameter $\gamma$ with the well-known
first-order Einsteinian effect (1.75 arcseconds at the Sun's
limb). Such a small value is beyond the presently available
possibilities; indeed, the Cassini test (Bertotti et al. 2003)
reached a $10^{-5}$ level, which has recently been questioned by
Kopeikin et al. (2006) who suggest a more realistic $10^{-4}$
error. Instead, it falls within the expected 0.02 microarcseconds
accuracy of the proposed LATOR mission (Turyshev et al. 2006),
which might be ready for launch in 2014. Also ASTROD (Ni 2002)
and, perhaps, GAIA (Vecchiato et al. 2003), could reach the
required sensitivity to measure such an effect. However, because
of technological and programmatic difficulties, the launch of an
ASTROD-like mission is not expected before 2025. GAIA is scheduled
to be launched in 2011
(http://gaia.esa.int/science-e/www/area/index.cfm?fareaid=26).

\section{Conclusions}
In this paper we have performed an independent test of the
hypothesis that the Pioneer anomaly can be explained by a
particular form of a recently proposed metric linear extension of
general relativity (Jaekel and Reynaud 2005a; 2005b) without
contradicting either the equivalence principle or the constraints
from planetary motions. Such a mechanism is based on a simple
explicit model involving, among other things, the occurrence of an
additional potential in the metric coefficient $g_{rr}$
parameterized in terms of  an adjustable free constant $\zeta_{\rm
P}$ and linearly varying with the distance $r$.

We have first compared the effects that the resulting
extra-acceleration, expressed in terms of the parameter
$\zeta_{\rm P}M$ set to the value which yields the magnitude of
the Pioneer anomalous acceleration, would also have on the Solar
System planetary motions with the latest available observational
determinations (Pitjeva 2005). It turns out that the determined
perihelion rates of the inner planets rule out the proposed
gravitational mechanism for explaining the Pioneer
extra-acceleration, at least in its present form. Conversely, the
determined extra-rates of the perihelia of the inner planets have
been used to measure $\zeta_{\rm P}M$ independently of the Pioneer
effect: it turns out that the so obtained value for such a
constant is three orders of magnitude smaller than it would be
required to reproduce the anomalous Pioneer acceleration. Another
independent test of the proposed model is represented, in
principle, by light deflection measurements in the proximity of
the Sun because $\zeta_{\rm P}$ also affects the propagation of
the electromagnetic waves. It turns out that the deflection angle
resulting from the value of $\zeta_{\rm P}M$ obtained  with the
perihelia extra-rates would be too small to be detected with the
present-day technology; future missions like LATOR, ASTROD and
GAIA will, instead, be able to measure such an effect.
%Conversely, the value for $\zeta_{\rm P}M$ which can be obtained
%from the determined perihelion rates is far from the one which
%reproduces the Pioneer effect.

The previous conclusions are based on the assumption that
$\zeta_{\rm P}$ is constant throughout the Solar System; Jaekel
and Reynaud, in fact, admit the possibility that it may vary with
the distance, but without further details.
%releasing any explicit
%detail or justification for such an hypothesis.

The results presented here further enforces the conclusions of
other studies (Iorio and Giudice 2006; Tangen 2006) pointing
towards an exclusion of a gravitational origin of the anomalous
features of motion experienced by the Pioneer spacecraft.

\section*{Acknowledgments}
I am grateful to S. Reynaud for the stimulating discussions had at
the Eleventh Marcel Grossmann Meeting on General Relativity held
in Berlin, 23-29 July 2006. Thanks also to E.V. Pitjeva for her
results about the outer planets of the Solar System.
%
%-----------------------------------------

\end{document}